\documentclass[twocolumn,superscriptaddress,amsmath,amssymb,aps,prc]{revtex4-2}
\usepackage{amsmath,amssymb}
\usepackage{graphicx}
\usepackage{dcolumn}
\usepackage{bm}
\usepackage{multirow}
\usepackage{textcomp}
\usepackage{url}
\usepackage{enumitem}
\usepackage{float}
\usepackage[dvipsnames]{xcolor}
\usepackage[colorlinks=true, allcolors=cyan]{hyperref}
\usepackage{orcidlink}

\begin{document}

\title{Chiral three-nucleon forces for the new local position-space two-nucleon potential in \textit{ab initio} many-body calculations}

\author{R. Z. Hu\,\orcidlink{0009-0002-8797-6622}}
\affiliation{School of Physics, and State Key Laboratory of Nuclear Physics and Technology, Peking University, Beijing 100871, China}
\author{J. G. Li\,\orcidlink{0000-0001-6002-7705}}
\affiliation{Institute of Modern Physics, Chinese Academy of Sciences, Lanzhou 730000, China}
\affiliation{Southern Center for Nuclear-Science Theory (SCNT), Institute of Modern Physics, Chinese Academy of Sciences, Huizhou 516000, China}
\author{S. Q. Fan\,\orcidlink{0000-0002-7400-9492}}
\affiliation{School of Physics, and State Key Laboratory of Nuclear Physics and Technology, Peking University, Beijing 100871, China}
\author{F. R. Xu\,\orcidlink{0000-0001-6699-0965}}\email[Corresponding author: ]{frxu@pku.edu.cn}
\affiliation{School of Physics, and State Key Laboratory of Nuclear Physics and Technology, Peking University, Beijing 100871, China}
\affiliation{Institute of Modern Physics, Chinese Academy of Sciences, Lanzhou 730000, China}
\affiliation{Southern Center for Nuclear-Science Theory (SCNT), Institute of Modern Physics, Chinese Academy of Sciences, Huizhou 516000, China}

\date{\today}

\begin{abstract}
Three-nucleon force (3NF) plays an important role in understanding the structure of finite nuclei and the saturation properties of infinite nuclear matter. More specifically, 3NF should be necessary for each two-nucleon force (2NF) to obtain more accurate description of nuclear systems. 3NF derived from the chiral effective field theory has been successful in \textit{ab initio} calculations of atomic nuclei. Most of established chiral nuclear forces have a nonlocal form in the momentum space. In this work, we construct a companion chiral 3NF specifically tailored to the new Idaho local position-space 2NF, and calculate binding energies and radii of nuclei up to $^{132}$Sn. We find that a chiral 3NF with hybrid local and nonlocal regulators has advantages in improving the nuclear structure calculations of both binding energies and radii with the new Idaho 2NF. The two low-energy constants of 3NF are constrained by the ground-state energies of $^3$H and $^{16}$O as suggested in a recent work. 
\end{abstract}

\maketitle

\section{Introduction}

One of the main goals of \textit{ab initio} nuclear theory is to understand atomic nuclei and nuclear matter from the fundamental degrees of freedom and interactions~\cite{Machleidt2023,10.3389/fphy.2023.1129094}. Within the framework of the chiral effective field theory ($\chi$EFT), two-nucleon and three-nucleon forces (2NF and 3NF, respectively) naturally emerge at different orders arranged by a proper power counting scheme~\cite{MACHLEIDT20111,RevModPhys.92.025004}. At a given chiral order, chiral nucleon interactions are renormalized by multiplying regulator functions that suppress high-momentum contributions beyond a certain cutoff momentum~\cite{PhysRevC.68.041001,EPELBAUM2005362}. Low-energy constants (LECs) appearing in chiral interactions are determined by available experimental data~\cite{PhysRevC.96.024004,RevModPhys.81.1773,PhysRevX.6.011019}. Chiral Hamiltonians obtained thus have been successfully applied to \textit{ab initio} calculations of strongly correlated many-body nuclei and nuclear matter~\cite{10.3389/fphy.2020.00379,MACHLEIDT2024104117,PhysRevLett.126.022501,Hu2022,PhysRevC.110.044317,PhysRevC.110.044316,PhysRevLett.132.182502,PhysRevLett.110.242501,Lee2020,PhysRevLett.106.192501,Elhatisari2015,Elhatisari2024,Gandolfi2020,Barrett2013}.

Nevertheless, open questions still remain for $\chi$EFT-based nuclear forces. One of the most important questions is whether the chiral interaction can simultaneously reproduce experimental ground-state energies and charge radii of medium-mass nuclei, and the saturation properties of nuclear matter~\cite{HEBELER20211,Machleidt2020,Elhatisari2024,RevModPhys.92.025004,PhysRevC.104.064312,BINDER2014119}.
Though many efforts have been made to address this issue by including many-body observables in the determination of 3NF LECs or by introducing different regularization schemes~\cite{PhysRevLett.110.192502,PhysRevC.93.011302,PhysRevC.96.014303,PhysRevC.102.054301,PhysRevC.101.014318,PhysRevC.91.054311,PhysRevC.86.054317,HUTHER2020135651,PhysRevLett.122.042501,PhysRevC.101.014318,PhysRevC.99.024313,PhysRevC.103.054001,PhysRevC.106.064002}, the situation is still unclear, and further investigations are needed~\cite{PhysRevC.102.034313,PhysRevC.102.044333,PhysRevC.109.L061302,PhysRevC.109.064003,PhysRevC.109.064314,PhysRevC.110.054322,PhysRevC.110.044003,PhysRevC.110.044004,tr4h-nl4d,kn79-f5m9,arXiv2401.06675,hu2025texas,hu2026fciqmc}.
While previous studies indicate that hybrid local-nonlocal (lnl) 3NF regulators can improve the simultaneous description of nuclear energies and radii compared with purely local regulators, a systematic underestimation of charge radii persists~\cite{PhysRevC.101.014318}.

Most of previous nuclear \textit{ab initio} calculations were based on nonlocal momentum-space potentials. Local position-space potentials were explored mainly for light-mass nuclei ($A\lesssim 16$) using quantum Monte Carlo (QMC) methods \cite{RevModPhys.87.1067,PhysRevLett.120.122502,PhysRevC.96.054007,PhysRevC.111.015801,PhysRevC.94.054007,PhysRevLett.120.052503,PhysRevLett.133.212501,PhysRevC.102.025501,PhysRevC.107.014314,PhysRevC.109.034005,PhysRevC.110.054316}. However, the situation for heavier nuclei is barely known due to the computation limits of QMC methods. In this work, we construct a companion chiral 3NF specifically tailored to the new Idaho local position-space 2NF~\cite{PhysRevC.107.034002}, and perform a systematic comparison between local and hybrid lnl 3NF regulators with the new family of 2NF plus 3NF. These should be useful for future \textit{ab initio} nuclear structure calculations.

\section{Chiral three-nucleon forces and regulators}
The $A$-body intrinsic Hamiltonian can be written as
\begin{equation}
    \hat{H}=\frac{1}{A} \sum_{i<j}^A \frac{(\boldsymbol{p}_i-\boldsymbol{p}_j)^2}{2 m}
    +\sum_{i<j}^A \hat{V}^{ij}_{\mathrm{NN}} +\sum_{i<j<k}^A \hat{V}^{ijk}_{\mathrm{3N}},
\end{equation}
where the first term denotes the intrinsic kinetic energy, whereas $\hat{V}_{\mathrm{NN}}$ and $\hat{V}_{\mathrm{3N}}$ indicate 2NF and 3NF, respectively.
In the present work, to improve many-body calculations, we want to construct a chiral 3NF which complements the chiral local position-space 2NF developed recently by the Idaho group~\cite{PhysRevC.107.034002}. The local position-space 2NF was obtained by the chiral expansion up to the N$^3$LO with a position-space regulator cutoff $R_\pi=1.2$ fm~\cite{PhysRevC.107.034002}. To speed up the convergences of many-body calculations of medium-mass nuclei, the 2NF is evolved to a low-momentum scale $\lambda=2.2$ fm$^{-1}$ using the  similarity renormalization group (SRG)~\cite{BOGNER201094,PhysRevC.75.061001}. The resulting low-resolution 2NF has been successfully applied to \textit{ab initio} no-core shell model (NCSM) calculations of low-lying states and electromagnetic properties of $^{10}$B~\cite{PhysRevLett.133.072502,PhysRevC.109.064316}.
We have checked that the $\lambda$ dependence of the results can be largely absorbed by adjusting the 3NF LECs, which equivalently considers induced-3NF effects.

The chiral 3NF at N$^2$LO consists of three topologies~\cite{PhysRevC.49.2932,PhysRevC.66.064001},
\begin{equation}
    \hat{V}_{\mathrm{3N}}= \hat{V}_{\mathrm{3N}}^{2\pi} + \hat{V}_{\mathrm{3N}}^{1\pi} + \hat{V}_{\mathrm{3N}}^{\mathrm{ct}}.
\end{equation}
The long-range two-$\pi$ exchange term  $\hat{V}_{\mathrm{3N}}^{2\pi}$ contains three pion-nucleon ($\pi N$) scattering LECs %To be consistent, 
which take $c_1=-0.74$ GeV$^{-1}$, $c_3=-3.61$ GeV$^{-1}$ and $c_4=2.44$ GeV$^{-1}$~\cite{PhysRevLett.115.192301,PhysRevC.107.034002,PhysRevC.103.054001} determined by the Roy-Steiner-equation scattering analysis at N$^2$LO~\cite{PhysRevLett.115.192301,HOFERICHTER20161}. 
The intermediate-range one-pion exchange $\hat{V}_{\mathrm{3N}}^{1\pi}$ and short-range three-nucleon contact term $\hat{V}_{\mathrm{3N}}^{\mathrm{ct}}$ contain two additional LECs, $c_{\text D}$ and $c_{\text E}$, respectively.

Similar to the 2NF, the chiral 3NF also needs to be applied with regulator functions to suppress high-momentum contributions. Several forms of the regulator function have been proposed, including the local form \cite{Navratil2007-yt}
\begin{equation}
    f_{\mathrm{local}} = \exp \Bigg[ -\left(\frac{|\boldsymbol{p}'_2-\boldsymbol{p}_2|^2}{\Lambda^2}\right)^2
    -\left(\frac{|\boldsymbol{p}'_3-\boldsymbol{p}_3|^2}{\Lambda^2}\right)^2\Bigg],
\end{equation}
the nonlocal form \cite{PhysRevC.66.064001}
\begin{equation}
    f_{\mathrm{nonlocal}} = \exp \Bigg[- \left(\frac{4p^2+3q^2}{4\Lambda^{'2}}\right)^2 - \left(\frac{4p'^2+3q'^2}{4\Lambda^{'2}}\right)^2 \Bigg],
\end{equation}
and the hybrid local-nonlocal (lnl) form \cite{PhysRevC.101.014318}
\begin{equation}
    f_{\mathrm{lnl}} = f_{\mathrm{local}} f_{\mathrm{nonlocal}},
    \label{lnl}
\end{equation}
with $\boldsymbol{p}_i$ ($\boldsymbol{p}'_i$) being the initial (final) momentum of $i$th nucleon, $p$ and $q$ ($p'$ and $q'$) being the magnitude of initial (final) Jacobi momenta. $\Lambda$ ($\Lambda^\prime$) is the $\chi$EFT hard cutoff. 
We have tested the two different regularization schemes to see the difference. The cutoffs are set to be $\Lambda = 500$ MeV for the local regulator, and $\Lambda = 700$ MeV and $\Lambda^\prime = 500$ MeV for the lnl regulator. Other different cutoffs of $\Lambda$ ($\Lambda^\prime$) have also been tested, obtaining similar results without conclusions changed.

\section{Parameterization of LECs \texorpdfstring{$c_{\text D}$}{} and \texorpdfstring{$c_{\text E}$}{} of chiral three-nucleon force}
There is still no unified agreement on strategies to constrain the 3NF LECs $c_{\text D}$ and $c_{\text E}$~\cite{Navratil2007-yt,PhysRevLett.110.192502,PhysRevC.103.054001,HUTHER2020135651}. However, at least two uncorrelated observables are required to determine $c_{\text D}$ and $c_{\text E}$ values. In this work, we use the $^3$H ground-state energy $E(^3\mathrm{H})$ as the first constraint on $c_{\text D}$ and $c_{\text E}$. Though there have been attempts to use the binding energy or charge radius of $^4$He as the second constraint~\cite{PhysRevLett.99.042501,PhysRevC.83.031301}, in many cases this strategy does not seem to be a good choice~\cite{HUTHER2020135651}, 
primarily due to the strong correlation between these observables~\cite{PhysRevC.104.064001}. The $\beta$-decay half-life of $^{3}$H may be used as a constraint~\cite{PhysRevLett.103.102502,PhysRevLett.107.062501,PhysRevC.67.055206,PhysRevLett.113.262504,PhysRevLett.96.232301,PhysRevC.98.044003,PhysRevC.104.064001}, because $c_{\text D}$ also appears in the short-range two-nucleon axial-vector current of the $\beta$ decay. In the recent work~\cite{HUTHER2020135651}, it was suggested to use the $^{16}$O ground-state energy $E(^{16}\mathrm{O})$ as a constraint which we follow in the present work.

\begin{figure}[t]
    \centering
    \includegraphics[width=0.37\textwidth]{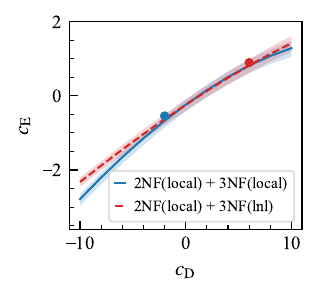}
    \caption{Relation between $c_{\text D}$ and $c_{\text E}$ obtained by the constraint of the $^3$H binding energy, for the local and lnl 3NF regulators. Shadows indicate the EFT uncertainty of 300 keV. The blue and red filled dots are the final optimized results by considering the $^{16}$O binding energy for 2NF(local)+3NF(local) and 2NF(local)+3NF(lnl), respectively.}
    \label{fig:cDcE}
\end{figure}

The $^3$H energy is calculated using the NCSM in the Jacobi coordinates~\cite{PhysRevC.61.044001}. The calculation is converged at a harmonic oscillator (HO) frequency around $\hbar\Omega=24$ MeV with  $N_{\mathrm{max}}=44$ HO shells. The $^3$H NCSM calculation with such large model space should be considered to be exact, and the many-body uncertainty should be ignored. The dominant uncertainty should be from the $\chi$EFT truncation.
The new Idaho 2NF is truncated at the N$^3$LO level~\cite{PhysRevC.107.034002}.
Uncertainties caused by $\chi$EFT can be estimated, e.g., for an observable $X$ at N$^2$LO and N$^3$LO, via \cite{Epelbaum2015,PhysRevC.93.044002,PhysRevC.98.014002}
\begin{equation}
\begin{aligned}
    \Delta X_{\mathrm{N^2LO}} =\max \Big( &Q^4|X_{\mathrm{LO}}|,\; Q^2|X_{\mathrm{LO}}-X_{\mathrm{NLO}}|,\\
    &Q|X_{\mathrm{NLO}}-X_{\mathrm{N^2LO}}|\Big),
\end{aligned}
\end{equation}
and 
\begin{equation}
\begin{aligned}
    \Delta X_{\mathrm{N^3LO}} =\max\Big( &Q^5|X_{\mathrm{LO}}|,\; Q^3|X_{\mathrm{LO}}-X_{\mathrm{NLO}}|,\\
    &Q^2|X_{\mathrm{NLO}}-X_{\mathrm{N^2LO}}|,\\
    &Q|X_{\mathrm{N^2LO}}-X_{\mathrm{N^3LO}}| \Big),
\end{aligned}
\end{equation}
respectively, where $Q$ is the EFT expansion scale measured by the ratio of the nucleon momentum transfer over the EFT hard cutoff $\Lambda$, which can be estimated by the pion mass over $\Lambda$. For $\Lambda\approx 600$ MeV, the typical value is $Q\approx 1/3$~\cite{PhysRevC.104.064001,HUTHER2020135651} which is used in quantifying the EFT uncertainty. We have estimated that uncertainties from the $\chi$EFT truncation at N$^2$LO and N$^3$LO are 446 keV and 149 keV, respectively, for the $^3$H ground-state energy.
Therefore, in optimizing the 3NF LECs $c_{\text D}$ and $c_{\text E}$, we do not use experimental data to strictly constrain the binding energy of $^3$H, whereas we allow the calculated energy to vary around the experimental value in a suitable range (approximately 300 keV) estimated with theoretical uncertainty. The constraint imposed by the $^3$H binding energy thus yields a correlation band between $c_{\text D}$ and $c_{\text E}$, as illustrated in Fig.~\ref{fig:cDcE}.

\begin{figure}[t]
    \centering
    \includegraphics[width=0.47\textwidth]{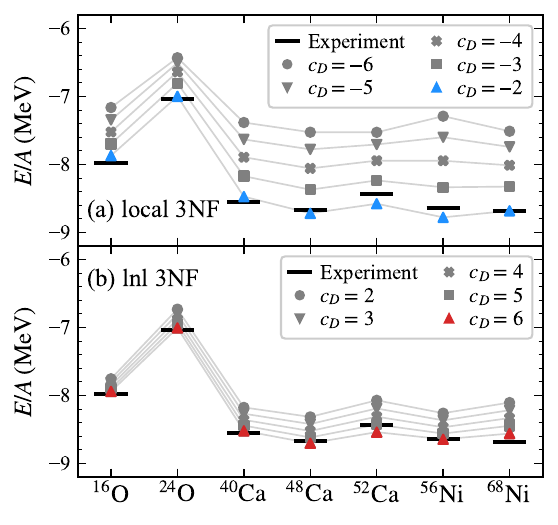}
    \caption{Ground-state energies of selected medium-mass doubly-closed-shell nuclei, calculated by  IMSRG with increasing the $c_{\text D}$ value in a step of 1.0, starting from $c_{\text D}=-6.0$ for the local 3NF regulator (a), and from $c_{\text D}=2.0$ for the lnl regulator (b). The black bars indicate experimental data~\cite{Wang_2021}.}
    \label{fig:fit_IMSRG}
\end{figure}

As suggested in Ref.~\cite{HUTHER2020135651}, the $^{16}$O ground-state energy is used as another constraint to fix the values of $c_{\text D}$ and $c_{\text E}$. We use the IMSRG in the Magnus formulation~\cite{PhysRevLett.106.222502,PhysRevC.92.034331} to calculate the $^{16}$O ground-state energy. The many-body Hamiltonian is normal ordered with respect to the reference state of the Hartree-Fock ground state as \cite{PhysRevLett.109.052501,ZHANG2022136958}
\begin{equation}
\begin{aligned}
    \hat{H} = &E_0 + \sum_{ij}f_{ij}\{ \hat{a}_i^\dagger \hat{a}_j \}
    + \frac{1}{2!^2} \sum_{ijkl}\Gamma_{ijkl}\{ \hat{a}_i^\dagger \hat{a}_j^\dagger \hat{a}_l \hat{a}_k\}\\
    &+ \frac{1}{3!^2} \sum_{ijklmn} W_{ijklmn} \{ \hat{a}_i^\dagger \hat{a}_j^\dagger \hat{a}_k^\dagger \hat{a}_n \hat{a}_m \hat{a}_l \},
\end{aligned}
\end{equation}
where $E_0$, $f$, $\Gamma$ and $W$ represent normal-ordered zero-, one-, two- and three-body terms, respectively. In the IMSRG evolution, operators are truncated at the two-body level. By comparisons with other many-body methods, it had been estimated that the many-body uncertainty from the IMSRG evolution with the two-body normal-ordered approximation is about 2\% in nuclear energy calculations~\cite{BINDER2014119,PhysRevLett.110.242501,PhysRevLett.109.052501,HUTHER2020135651,PhysRevC.110.044317,PhysRevC.110.044316}.
We have carefully calculated the ground-state energies of medium-mass nuclei at closed shells by gradually increasing the $c_{\text D}$ value with $c_{\text E}$ also changing according to the relation obtained by the $E(^3\mathrm{H})$ constraint. It is found that the calculated energies are monotonously lowered with increasing the $c_{\text D}$ value. For the local 3NF regulator, calculations with $c_{\text D}$ increasing starting from $-6.0$ in a step of 1.0 show that $c_{\text D}=-2.0$ gives good descriptions of experimental ground-state energies of the medium-mass closed-shell nuclei, as shown in the upper panel of Fig.~\ref{fig:fit_IMSRG}. For the lnl 3NF regulator, calculations with $c_{\text D}$ increasing starting from $2.0$ in a step of 1.0 show that $c_{\text D}=6.0$ gives good descriptions of ground-state energies of the nuclei, see the lower panel of Fig.~\ref{fig:fit_IMSRG}.

The optimal $c_{\text D}$ and $c_{\text E}$ values are summarized in Table~\ref{table:3N_LECs} for the constructed 3NF with a local or lnl regulator. Note that we have only selected the integer values of $c_{\text D}$ with a step of 1.0 in fitting, since errors (uncertainties) originating from IMSRG many-body and $\chi$EFT truncations are larger than the $E(^{16}{\text O})$ change caused by a change of one unit in the $c_{\text D}$ value. This strategy is similar to that used in Ref.~\cite{HUTHER2020135651}.

\begin{table}[t]
    \centering
    \caption{Optimal $c_{\text D}$ and $c_{\text E}$ values, and calculated ground-state energies (in MeV) of $^{3}$H and $^{16}$O, compared with data~\cite{Wang_2021}. The new local position-space Idaho potential~\cite{PhysRevC.107.034002} is used for the 2NF, while we construct a local or local-nonlocal (lnl) 3NF for this 2NF.}
    \renewcommand{\arraystretch}{1.45}
    \setlength{\tabcolsep}{1.0mm}
    \begin{tabular}{ccccc}
    \hline\hline
    & $c_{\text D}$ & $c_{\text E}$ & $E$($^3$H) & $E$($^{16}$O) \\
    \hline
    3NF(local) & $-2.0$ & $-0.541$ & $-8.78$  & $-126.01$ \\
    3NF(lnl) & $6.0$ & $0.894$ & $-8.73$ & $-127.15$ \\
    Expt. & & & $-8.482$ & $-127.619$ \\
    \hline\hline
\end{tabular}
\label{table:3N_LECs}
\end{table}

\section{Applications to the calculations of nuclear energies and radii}

\begin{figure*}[t]
    \centering
    \includegraphics[width=0.95\textwidth]{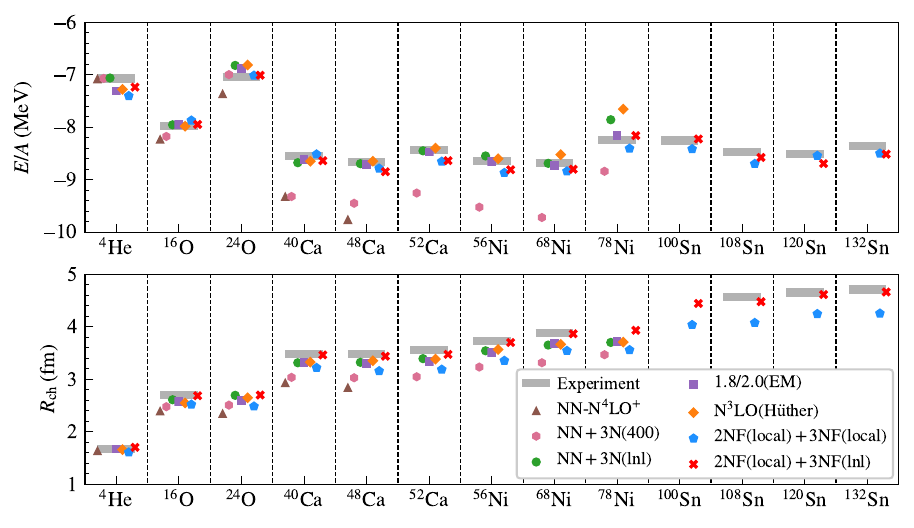}
    \caption{Ground-state energies per nucleon and charge radii for selected doubly-closed-shell nuclei from $^{4}$He to $^{132}$Sn.
   The present calculations using our local and local-nonlocal 3NFs are labeled by 2NF(local)+3NF(local) and 2NF(local)+3NF(lnl), respectively. They are compared with other 2NF-only calculations by the NN-N$^4$LO$^+$ interaction~\cite{PhysRevC.106.064002}, and 3NF-included calculations by NN+3N(400)~\cite{PhysRevLett.109.052501,PhysRevC.89.061301}, NN+3N(lnl)~\cite{PhysRevC.101.014318}, 1.8/2.0(EM) \cite{PhysRevC.83.031301,PhysRevLett.126.022501,PhysRevC.96.014303} and N$^3$LO(H\"uther)~\cite{HUTHER2020135651}.     Experimental data are taken from Refs.~\cite{Wang_2021,ANGELI201369,GarciaRuiz2016,PhysRevLett.129.132501}.}
    \label{fig:properties_comparision}
\end{figure*}

The constructed 3NF forms a family with the new local position-space 2NF proposed by the Idaho group~\cite{PhysRevC.107.034002}. We have used the new family of two- plus three-nucleon interactions to calculate the binding energies and charge radii of nuclei over a large range from $^4$He to $^{132}$Sn.

\subsection{Closed-shell nuclei}

For closed-shell nuclei, the single-reference IMSRG can be used to calculate energies and radii of the ground states. The model space is restricted by the single-particle basis truncation $e=2n+l\leq e_{\mathrm{max}}$ and the 3NF matrix element truncation $e_1+e_2+e_3\leq E_{3\mathrm{max}}$. In this work, we use $e_{\mathrm{max}}=14$ and $E_{3\mathrm{max}}=24$ \cite{Miyagi2023} with optimized HO frequencies ($\hbar\Omega$) to ensure model-space convergence in all cases.

We compare the present calculations with those obtained using some other well-established interactions, namely NN-N$^4$LO$^+$ \cite{PhysRevC.106.064002} (with $\Lambda=500$ MeV), NN+3N(400)~\cite{PhysRevLett.109.052501,PhysRevC.89.061301}, NN+3N(lnl) \cite{PhysRevC.101.014318}, 1.8/2.0(EM) \cite{PhysRevC.83.031301,PhysRevLett.126.022501,PhysRevC.96.014303} and N$^3$LO(H\"uther) (both 2NF and 3NF at N$^3$LO with $\Lambda=500$ MeV)~\cite{HUTHER2020135651}, shown in Fig.~\ref{fig:properties_comparision}.
We see that the present 2NF(local) plus 3NF(local) family describes binding energies well, but underestimates charge radii in general. If we use the hybrid 3NF regulator of locality and nonlocality defined as Eq.~(\ref{lnl}), the family of 2NF(local) plus 3NF(lnl) gives the most accurate description of both binding energies and charge radii for the closed-shell nuclei from $^4$He to $^{132}$Sn, with deviations from experimental data below 2\%, as shown in Fig.~\ref{fig:properties_comparision}. The lnl 3NF regulator scheme was suggested in Ref.~\cite{PhysRevC.101.014318}, and tested with the family of a momentum-space N$^3$LO 2NF~\cite{PhysRevC.68.041001} plus the lnl N$^2$LO 3NF~\cite{PhysRevC.101.014318}. While the lnl 3NF regulator significantly improves the description of nuclear radii, calculated charge radii remain considerably smaller than experimental values~\cite{PhysRevC.101.014318}. Compared with this family based on the momentum-space 2NF plus the lnl 3NF [labeled NN+3N(lnl) in Fig.~\ref{fig:properties_comparision}], the present family utilizes the new position-space 2NF plus the lnl 3NF, but adopts a different fitting strategy to determine $c_\mathrm{D}$ and $c_\mathrm{E}$. As shown in Fig.~\ref{fig:properties_comparision} with comparisons with other potentials, the present calculations with the new family of the local position-space N$^3$LO 2NF plus the lnl N$^2$LO 3NF [labeled 2NF(local)+3NF(lnl) in Fig.~\ref{fig:properties_comparision}] gives the best description of both energies and radii, clearly better than calculations with the family of 2NF(local)+3NF(local).

\subsection{Open-shell nuclei}
\begin{figure*}[htbp]
    \centering
    \includegraphics[width=0.95\textwidth]{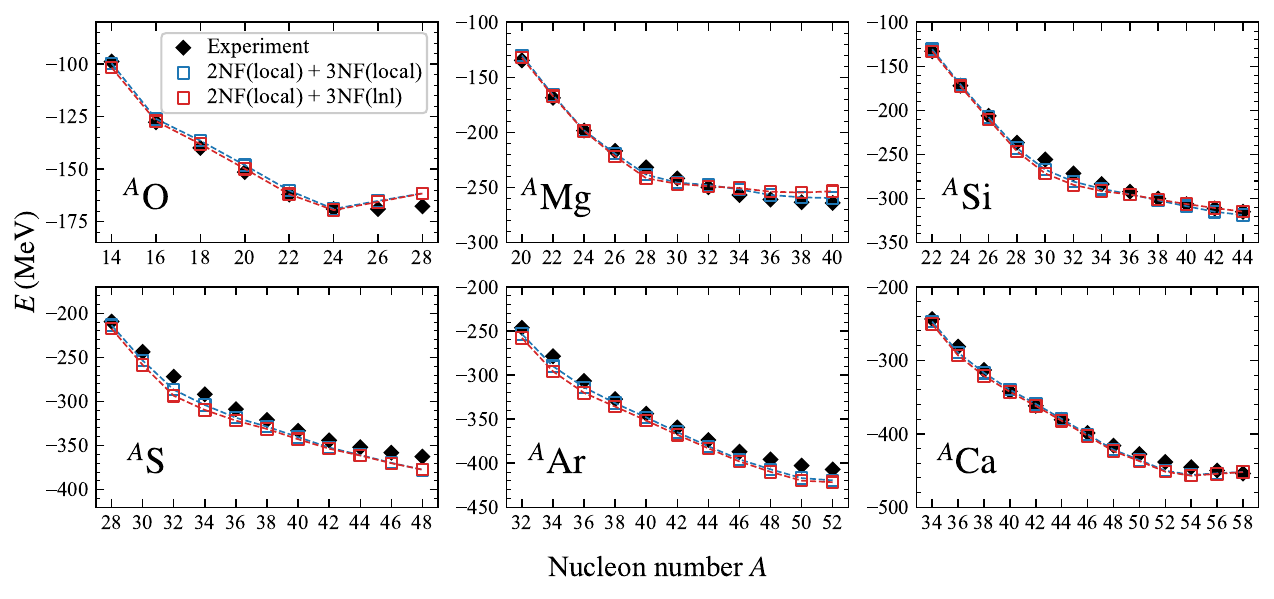}
    \caption{Ground-state energies of O, Mg, Si, S, Ar and Ca isotopes, calculated in this work using the new families of the present 3NF (local or local-nonlocal) with the new Idaho position-space 2NF (local) within the many-body VS-IMSRG. Experimental data are taken from the 2020 atomic mass evaluation (AME2020)~\cite{Wang_2021}.}
    \label{fig:chain_all_energy}
\end{figure*}

\begin{figure*}[htbp]
    \centering
    \includegraphics[width=0.95\textwidth]{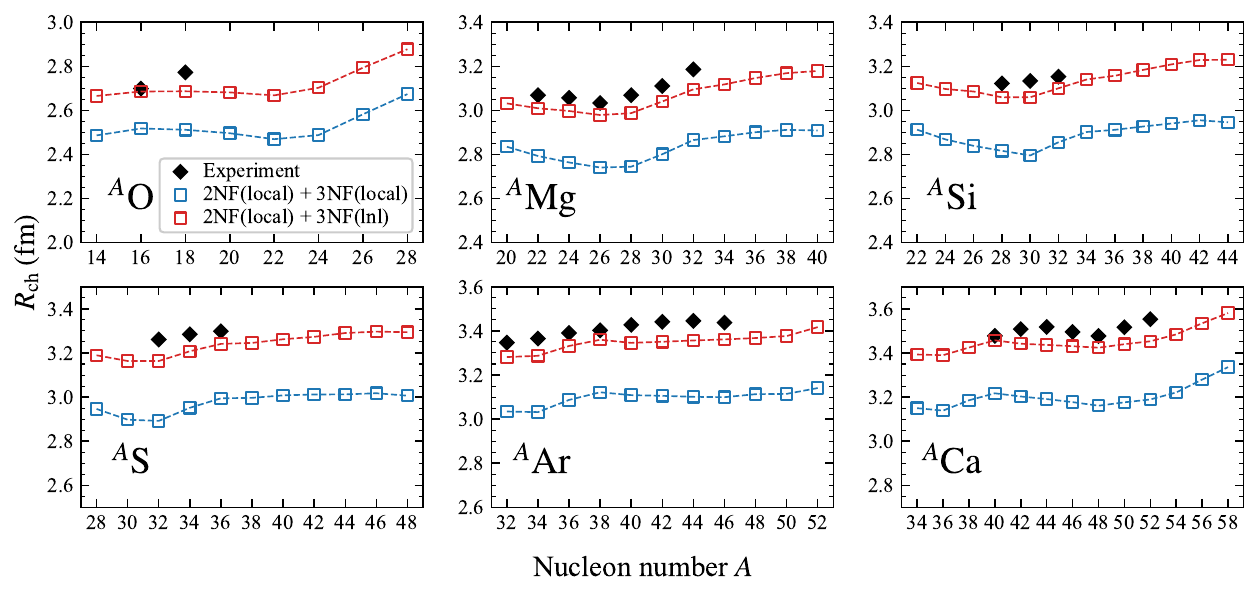}
    \caption{Similar to Fig.~\ref{fig:chain_all_energy} but for charge radii. Experimental charge radii are taken from Ref.~\cite{ANGELI201369,GarciaRuiz2016,PhysRevLett.129.132501}.}
    \label{fig:chain_all_radius}
\end{figure*}

Open-shell nuclei can be calculated using the so-called valence-space IMSRG (VS-IMSRG) in which the valence-space effective Hamiltonian is obtained by the VS-IMSRG evolution~\cite{PhysRevLett.106.222502}. Furthermore, to reduce the residual 3NF effect and choose a more appropriate shell-model core, the nucleus-dependent VS-IMSRG Hamiltonian with fractional filling of open-shell orbitals is used, named the ensemble normal ordering (ENO)~\cite{PhysRevLett.118.032502}. The ENO VS-IMSRG evolution is also used to derive valence-space effective operators of other observables. The valence-space Hamiltonian is then diagonalized using the parallel shell model code \texttt{kshell}~\cite{SHIMIZU2019372}. In the present VS-IMSRG calculation, the valence space is defined as follows: $p$ shell with $^4$He core for ${}^{14}\mathrm{O}$; $sd$ shell with $^{16}$O core for ${}^{17{\text -}28}\mathrm{O}$, ${}^{20{\text -}32}\mathrm{Mg}$, ${}^{22{\text -}34}\mathrm{Si}$, ${}^{28{\text -}36}\mathrm{S}$, ${}^{32{\text -}38}\mathrm{Ar}$ and ${}^{34{\text -}38}\mathrm{Ca}$; proton $sd$ and neutron $pf$ shells with ${}^{28}\mathrm{O}$ core for ${}^{34{\text -}40}\mathrm{Mg}$, ${}^{36{\text -}44}\mathrm{Si}$, ${}^{38{\text -}48}\mathrm{S}$ and ${}^{40{\text -}52}\mathrm{Ar}$; and $pf$ shell with ${}^{40}\mathrm{Ca}$ core for ${}^{41{\text -}58}\mathrm{Ca}$.

Figure~\ref{fig:chain_all_energy} shows the systematics of calculated ground-state energies, compared with experimental data, for oxygen, magnesium, silicon, sulfur, argon and calcium isotopic chains. We find that both local and lnl 3NFs give overall satisfying agreements with experimental ground-state energies for all isotopic chains investigated. 
Figure~\ref{fig:chain_all_radius} shows charge radii for the isotopic chains studied. The situation is similar to closed-shell cases. The local-nonlocal N$^2$LO 3NF connected to the new local position-space N$^3$LO 2NF provides good description of charge radii, while local 3NF with the same position-space 2NF underestimates the radii systematically. The change from a local 3NF regulator to a local-nonlocal 3NF regulator brings about $7\%$ increase in the charge radius for medium-mass nuclei. The increase was also observed in previous studies~\cite{PhysRevC.101.014318,Soma2021-lv}.

\subsection{Charge density distributions}

Nuclear charge density distribution can provide more detailed information of nuclear structure, offering an even finer test for chiral Hamiltonians~\cite{Lu2019,PhysRevC.91.051301,PhysRevC.101.014318}. Such investigations have been carried out for decades also in \textit{ab initio} calculations using a variety of many-body approaches and interactions~\cite{PhysRevC.95.034319,Soma2021-lv,PhysRevLett.119.222505,PhysRevLett.125.182501,PhysRevC.62.014001,Hagen2015,PhysRevC.89.024305,PhysRevLett.113.192501,PhysRevC.96.054007,PhysRevC.111.024314}. Theoretically, the nuclear charge density can be calculated by folding the point-nucleon density with the intrinsic form factor of the free nucleons expressed in terms of the Sachs and Pauli form factors, see Refs.~\cite{PhysRevC.1.1260,Soma2021-lv,PhysRevC.103.054310} for details of the calculation.

\begin{figure}[t]
    \centering
    \includegraphics[width=0.45\textwidth]{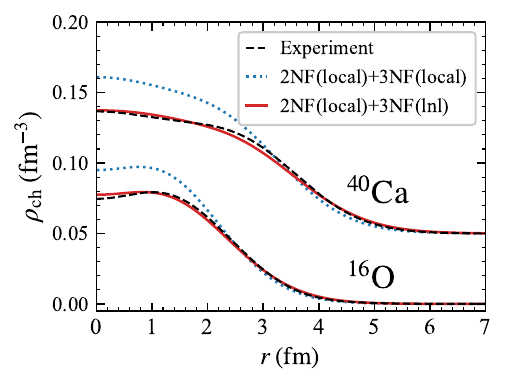}
    \caption{Charge density distributions of $^{16}$O and $^{40}$Ca, calculated with 2NF(local)+3NF(local) and 2NF(local)+3NF(lnl), compared with experimental data~\cite{DEVRIES1987495}. The density of $^{40}$Ca is shifted by 0.05 fm$^{-3}$ for better readability.}
    \label{fig:density}
\end{figure}

In this work, we focus on the charge densities of two representative closed-shell nuclei, ${}^{16}\mathrm{O}$ and ${}^{40}\mathrm{Ca}$, whose experimental charge densities have been well measured. The converged calculations are performed using the single-reference IMSRG with $e_{\mathrm{max}}=14$, $E_{3\mathrm{max}}=24$ and an optimal HO frequency of $\hbar \Omega=16$ MeV. In both nuclei, charge density distributions calculated using 2NF(local)+3NF(lnl) are in excellent agreements with experimental data, while the densities obtained by 2NF(local)+3NF(local) are significantly larger than data in the central region of the nuclei, as shown in Fig.~\ref{fig:density}. The poorer description of the interior charge density obtained with the local 3NF may be attributed to the fact that local regulators are typically harder than their nonlocal counterparts, which may induce stronger short-range correlations~\cite{HEBELER20211}. The hybrid lnl regulator, by incorporating an additional nonlocal Gaussian factor, softens the short-range correlations, which results in a decrease in the interior density and an increase in the exterior density, thereby leading to an increased radius. Experimental measurements of nuclear charge density distribution remain a frontier area of nuclear physics studies~\cite{10.1093/ptep/pts043,PhysRevLett.118.262501,ANTONOV201160}. Theoretical calculations of density distribution based on the first principles can provide deep insights into nucleon-nucleon interactions.

\section{Summary}
Understanding the properties of atomic nuclei from first principles represents a central challenge in nuclear physics. Within the framework of chiral effective field theory, nuclear forces are hierarchically organized, where three-nucleon forces emerge as a crucial component for accurate descriptions of nuclear structure. While significant progress has been made, it remains a persistent challenge to develop a unified chiral Hamiltonian that accurately describes both binding energies and charge radii across the nuclear chart.

In this work, we have constructed chiral three-nucleon forces at N$^2$LO , which form new families with the new high-quality local position-space two-nucleon interaction established by the Idaho group. The two low-energy constants, $c_{\text D}$ and $c_{\text E}$, of the 3NF are constrained using the ground-state energies of $^3$H and $^{16}$O. We find that employing a local-nonlocal hybrid regularization scheme for the 3NF is particularly effective. The resulted chiral Hamiltonian is applied to nuclear many-body calculations, and provides the excellent description of both ground-state energies and charge radii for a wide range of nuclei from $^4$He up to $^{132}$Sn. This work highlights that the regularization scheme of the chiral 3NF can sensitively influence the quality of many-body predictions, especially for nuclear radii. The chiral Hamiltonian developed in this work should be useful for future high-precision \textit{ab initio} calculations of nuclei.

\begin{acknowledgments}
The IMSRG calculations were performed using the \texttt{imsrg++} code \cite{imsrgcode}, and chiral 3NF matrix elements were generated using the \texttt{NuHamil} code \cite{Miyagi2023}.
This work has been supported by the National Key R\&D Program of China under Grants No. 2024YFA1610900, 2023YFA1606400; the National Natural Science Foundation of China under Grants No. 12335007, 12535008, 12441506. We acknowledge the High-Performance Computing Platform of Peking University for providing computational resources.
\end{acknowledgments}

\section*{Data Availability}
The data that support the findings of this article are openly available~\cite{Data}.

\bibliographystyle{modified-apsrev4-2.bst}
\bibliography{reference}

@PREAMBLE{
"\providecommand{\noopsort}[1]{}"
# "\providecommand{\singleletter}[1]{#1}%"
}

@Article{Machleidt2023,
author={Machleidt, R.},
title={What is \textit{ab initio}?},
journal={Few-Body Syst.},
year={2023},
month={Oct},
day={03},
volume={64},
number={4},
pages={77},
issn={1432-5411},
doi={10.1007/s00601-023-01857-2},
url={https://doi.org/10.1007/s00601-023-01857-2}
}

@ARTICLE{10.3389/fphy.2023.1129094,
title     = {What is \textit{ab initio} in nuclear theory?},
  volume    = {11},
  issn      = {2296-424X},
  pages={1129094},
  url       = {http://dx.doi.org/10.3389/fphy.2023.1129094},
  doi       = {10.3389/fphy.2023.1129094},
  journal   = {Front. Phys.},
  publisher = {Frontiers Media SA},
  author    = {Ekstr\"{o}m,  A. and Forss\'en,  C. and Hagen,  G. and Jansen,  G. R. and Jiang,  W. and Papenbrock,  T.},
  year      = {2023},
  month     = {Feb}
}

@article{MACHLEIDT2024104117,
title = {Recent advances in chiral EFT based nuclear forces and their applications},
journal = {Prog. Part. Nucl. Phys.},
volume = {137},
pages = {104117},
year = {2024},
issn = {0146-6410},
doi = {https://doi.org/10.1016/j.ppnp.2024.104117},
url = {https://www.sciencedirect.com/science/article/pii/S0146641024000218},
author = {R. Machleidt and F. Sammarruca}
}

@article{RevModPhys.81.1773,
  title = {Modern theory of nuclear forces},
  author = {Epelbaum, E. and Hammer, H.-W. and Mei\ss{}ner, Ulf-G.},
  journal = {Rev. Mod. Phys.},
  volume = {81},
  issue = {4},
  pages = {1773--1825},
  numpages = {0},
  year = {2009},
  month = {Dec},
  publisher = {American Physical Society},
  doi = {10.1103/RevModPhys.81.1773},
  url = {https://link.aps.org/doi/10.1103/RevModPhys.81.1773}
}

@article{EPELBAUM2005362,
title = {The two-nucleon system at next-to-next-to-next-to-leading order},
journal = {Nucl. Phys. A},
volume = {747},
number = {2},
pages = {362-424},
year = {2005},
issn = {0375-9474},
doi = {https://doi.org/10.1016/j.nuclphysa.2004.09.107},
url = {https://www.sciencedirect.com/science/article/pii/S0375947404010747},
author = {E. Epelbaum and W. Glöckle and Ulf-G. Meißner}
}

@article{PhysRevC.96.024004,
  title = {High-quality two-nucleon potentials up to fifth order of the chiral expansion},
  author = {Entem, D. R. and Machleidt, R. and Nosyk, Y.},
  journal = {Phys. Rev. C},
  volume = {96},
  issue = {2},
  pages = {024004},
  numpages = {19},
  year = {2017},
  month = {Aug},
  publisher = {American Physical Society},
  doi = {10.1103/PhysRevC.96.024004},
  url = {https://link.aps.org/doi/10.1103/PhysRevC.96.024004}
}

@article{PhysRevC.68.041001,
  title = {Accurate charge-dependent nucleon-nucleon potential at fourth order of chiral perturbation theory},
  author = {Entem, D. R. and Machleidt, R.},
  journal = {Phys. Rev. C},
  volume = {68},
  issue = {4},
  pages = {041001},
  numpages = {5},
  year = {2003},
  month = {Oct},
  publisher = {American Physical Society},
  doi = {10.1103/PhysRevC.68.041001},
  url = {https://link.aps.org/doi/10.1103/PhysRevC.68.041001}
}

@article{RevModPhys.92.025004,
  title = {Nuclear effective field theory: Status and perspectives},
  author = {Hammer, H.-W. and K\"onig, Sebastian and van Kolck, U.},
  journal = {Rev. Mod. Phys.},
  volume = {92},
  issue = {2},
  pages = {025004},
  numpages = {66},
  year = {2020},
  month = {Jun},
  publisher = {American Physical Society},
  doi = {10.1103/RevModPhys.92.025004},
  url = {https://link.aps.org/doi/10.1103/RevModPhys.92.025004}
}

@article{PhysRevC.107.034002,
  title = {Local position-space two-nucleon potentials from leading to fourth order of chiral effective field theory},
  author = {Saha, S. K. and Entem, D. R. and Machleidt, R. and Nosyk, Y.},
  journal = {Phys. Rev. C},
  volume = {107},
  issue = {3},
  pages = {034002},
  numpages = {27},
  year = {2023},
  month = {Mar},
  publisher = {American Physical Society},
  doi = {10.1103/PhysRevC.107.034002},
  url = {https://link.aps.org/doi/10.1103/PhysRevC.107.034002}
}

@article{10.3389/fphy.2020.00379,
  title     = {A Guided Tour of \textit{ab initio} Nuclear Many-Body Theory},
  volume    = {8},
  issn      = {2296-424X},
  url       = {http://dx.doi.org/10.3389/fphy.2020.00379},
  doi       = {10.3389/fphy.2020.00379},
  journal   = {Front. Phys.},
  publisher = {Frontiers Media SA},
  author    = {Hergert,  Heiko},
  year      = {2020},
  month     = {Oct},
  pages     = {379}
}

@article{HEBELER20211,
title = {Three-nucleon forces: Implementation and applications to atomic nuclei and dense matter},
journal = {Phys. Rep.},
volume = {890},
pages = {1-116},
year = {2021},
issn = {0370-1573},
doi = {https://doi.org/10.1016/j.physrep.2020.08.009},
url = {https://www.sciencedirect.com/science/article/pii/S0370157320303409},
author = {Kai Hebeler}
}

@article{MACHLEIDT20111,
title = {Chiral effective field theory and nuclear forces},
journal = {Phys. Rep.},
volume = {503},
number = {1},
pages = {1-75},
year = {2011},
issn = {0370-1573},
doi = {https://doi.org/10.1016/j.physrep.2011.02.001},
url = {https://www.sciencedirect.com/science/article/pii/S0370157311000457},
author = {R. Machleidt and D.R. Entem}
}

@article{PhysRevC.103.054001,
  title = {Light nuclei with semilocal momentum-space regularized chiral interactions up to third order},
  author = {Maris, P. and others},
  collaboration = {LENPIC Collaboration},
  journal = {Phys. Rev. C},
  volume = {103},
  issue = {5},
  pages = {054001},
  numpages = {20},
  year = {2021},
  month = {May},
  publisher = {American Physical Society},
  doi = {10.1103/PhysRevC.103.054001},
  url = {https://link.aps.org/doi/10.1103/PhysRevC.103.054001}
}

@article{RevModPhys.87.1067,
  title = {Quantum Monte Carlo methods for nuclear physics},
  author = {Carlson, J. and others},
  journal = {Rev. Mod. Phys.},
  volume = {87},
  issue = {3},
  pages = {1067--1118},
  numpages = {52},
  year = {2015},
  month = {Sep},
  publisher = {American Physical Society},
  doi = {10.1103/RevModPhys.87.1067},
  url = {https://link.aps.org/doi/10.1103/RevModPhys.87.1067}
}

@article{PhysRevLett.110.192502,
  title = {Optimized Chiral Nucleon-Nucleon Interaction at Next-to-Next-to-Leading Order},
  author = {Ekstr\"om, A. and others},
  journal = {Phys. Rev. Lett.},
  volume = {110},
  issue = {19},
  pages = {192502},
  numpages = {5},
  year = {2013},
  month = {May},
  publisher = {American Physical Society},
  doi = {10.1103/PhysRevLett.110.192502},
  url = {https://link.aps.org/doi/10.1103/PhysRevLett.110.192502}
}

@article{PhysRevC.102.054301,
  title = {Accurate bulk properties of nuclei from $A=2$ to $\ensuremath{\infty}$ from potentials with $\mathrm{\ensuremath{\Delta}}$ isobars},
  author = {Jiang, W. G. and others},
  journal = {Phys. Rev. C},
  volume = {102},
  issue = {5},
  pages = {054301},
  numpages = {8},
  year = {2020},
  month = {Nov},
  publisher = {American Physical Society},
  doi = {10.1103/PhysRevC.102.054301},
  url = {https://link.aps.org/doi/10.1103/PhysRevC.102.054301}
}

@article{PhysRevC.91.051301,
  title = {Accurate nuclear radii and binding energies from a chiral interaction},
  author = {Ekstr\"om, A. and others},
  journal = {Phys. Rev. C},
  volume = {91},
  issue = {5},
  pages = {051301},
  numpages = {7},
  year = {2015},
  month = {May},
  publisher = {American Physical Society},
  doi = {10.1103/PhysRevC.91.051301},
  url = {https://link.aps.org/doi/10.1103/PhysRevC.91.051301}
}

@article{PhysRevC.49.2932,
  title = {Few-nucleon forces from chiral Lagrangians},
  author = {van Kolck, U.},
  journal = {Phys. Rev. C},
  volume = {49},
  issue = {6},
  pages = {2932--2941},
  numpages = {0},
  year = {1994},
  month = {Jun},
  publisher = {American Physical Society},
  doi = {10.1103/PhysRevC.49.2932},
  url = {https://link.aps.org/doi/10.1103/PhysRevC.49.2932}
}

@article{PhysRevC.66.064001,
  title = {Three-nucleon forces from chiral effective field theory},
  author = {Epelbaum, E. and others},
  journal = {Phys. Rev. C},
  volume = {66},
  issue = {6},
  pages = {064001},
  numpages = {17},
  year = {2002},
  month = {Dec},
  publisher = {American Physical Society},
  doi = {10.1103/PhysRevC.66.064001},
  url = {https://link.aps.org/doi/10.1103/PhysRevC.66.064001}
}

@Article{Hu2022,
author={Hu, Baishan
and others},
title={\textit{Ab initio} predictions link the neutron skin of $^{208}$Pb to nuclear forces},
journal={Nat. Phys.},
year={2022},
month={Oct},
day={01},
volume={18},
number={10},
pages={1196-1200},
issn={1745-2481},
doi={10.1038/s41567-022-01715-8},
url={https://doi.org/10.1038/s41567-022-01715-8}
}

@article{PhysRevLett.113.192501,
  title = {Quantum Monte Carlo Calculations of Light Nuclei Using Chiral Potentials},
  author = {Lynn, J. E. and Carlson, J. and Epelbaum, E. and Gandolfi, S. and Gezerlis, A. and Schwenk, A.},
  journal = {Phys. Rev. Lett.},
  volume = {113},
  issue = {19},
  pages = {192501},
  numpages = {5},
  year = {2014},
  month = {Nov},
  publisher = {American Physical Society},
  doi = {10.1103/PhysRevLett.113.192501},
  url = {https://link.aps.org/doi/10.1103/PhysRevLett.113.192501}
}

@article{PhysRevC.96.054007,
  title = {Quantum Monte Carlo calculations of light nuclei with local chiral two- and three-nucleon interactions},
  author = {Lynn, J. E. and Tews, I. and Carlson, J. and Gandolfi, S. and Gezerlis, A. and Schmidt, K. E. and Schwenk, A.},
  journal = {Phys. Rev. C},
  volume = {96},
  issue = {5},
  pages = {054007},
  numpages = {19},
  year = {2017},
  month = {Nov},
  publisher = {American Physical Society},
  doi = {10.1103/PhysRevC.96.054007},
  url = {https://link.aps.org/doi/10.1103/PhysRevC.96.054007}
}

@article{PhysRevC.111.024314,
  title = {Static and dynamic properties of atomic nuclei with high-resolution potentials},
  author = {Gnech, Alex and Lovato, Alessandro and Rocco, Noemi},
  journal = {Phys. Rev. C},
  volume = {111},
  issue = {2},
  pages = {024314},
  numpages = {13},
  year = {2025},
  month = {Feb},
  publisher = {American Physical Society},
  doi = {10.1103/PhysRevC.111.024314},
  url = {https://link.aps.org/doi/10.1103/PhysRevC.111.024314}
}

@article{Navratil2007-yt,
  title = {Local three-nucleon interaction from chiral effective field theory},
  volume = {41},
  ISSN = {1432-5411},
  url = {http://dx.doi.org/10.1007/s00601-007-0193-3},
  DOI = {10.1007/s00601-007-0193-3},
  number = {3-4},
  journal = {Few-Body Systems},
  publisher = {Springer Science and Business Media LLC},
  author = {Navrátil,  P.},
  year = {2007},
  month = {Nov},
  pages = {117–140}
}

@article{PhysRevC.89.061301,
  title = {Chiral two- and three-nucleon forces along medium-mass isotope chains},
  author = {Som\`a, V. and Cipollone, A. and Barbieri, C. and Navr\'atil, P. and Duguet, T.},
  journal = {Phys. Rev. C},
  volume = {89},
  issue = {6},
  pages = {061301},
  numpages = {5},
  year = {2014},
  month = {Jun},
  publisher = {American Physical Society},
  doi = {10.1103/PhysRevC.89.061301},
  url = {https://link.aps.org/doi/10.1103/PhysRevC.89.061301}
}

@article{BINDER2014119,
title = {\textit{Ab initio} path to heavy nuclei},
journal = {Phys. Lett. B},
volume = {736},
pages = {119-123},
year = {2014},
issn = {0370-2693},
doi = {https://doi.org/10.1016/j.physletb.2014.07.010},
url = {https://www.sciencedirect.com/science/article/pii/S0370269314004961},
author = {Sven Binder and Joachim Langhammer and Angelo Calci and Robert Roth}
}

@article{PhysRevLett.99.042501,
  title = {Structure of $A=10-13$ Nuclei with Two- Plus Three-Nucleon Interactions from Chiral Effective Field Theory},
  author = {Navr\'atil, P. and Gueorguiev, V. G. and Vary, J. P. and Ormand, W. E. and Nogga, A.},
  journal = {Phys. Rev. Lett.},
  volume = {99},
  issue = {4},
  pages = {042501},
  numpages = {4},
  year = {2007},
  month = {Jul},
  publisher = {American Physical Society},
  doi = {10.1103/PhysRevLett.99.042501},
  url = {https://link.aps.org/doi/10.1103/PhysRevLett.99.042501}
}

@article{PhysRevC.96.014303,
  title = {Saturation with chiral interactions and consequences for finite nuclei},
  author = {Simonis, J. and Stroberg, S. R. and Hebeler, K. and Holt, J. D. and Schwenk, A.},
  journal = {Phys. Rev. C},
  volume = {96},
  issue = {1},
  pages = {014303},
  numpages = {11},
  year = {2017},
  month = {Jul},
  publisher = {American Physical Society},
  doi = {10.1103/PhysRevC.96.014303},
  url = {https://link.aps.org/doi/10.1103/PhysRevC.96.014303}
}

@article{PhysRevLett.126.022501,
  title = {\textit{Ab Initio} Limits of Atomic Nuclei},
  author = {Stroberg, S. R. and Holt, J. D. and Schwenk, A. and Simonis, J.},
  journal = {Phys. Rev. Lett.},
  volume = {126},
  issue = {2},
  pages = {022501},
  numpages = {6},
  year = {2021},
  month = {Jan},
  publisher = {American Physical Society},
  doi = {10.1103/PhysRevLett.126.022501},
  url = {https://link.aps.org/doi/10.1103/PhysRevLett.126.022501}
}

@article{Lee2020,
  title = {Recent Progress in Nuclear Lattice Simulations},
  volume = {8},
  ISSN = {2296-424X},
  url = {http://dx.doi.org/10.3389/fphy.2020.00174},
  DOI = {10.3389/fphy.2020.00174},
  journal = {Front. Phys.},
  publisher = {Frontiers Media SA},
  author = {Lee,  Dean},
  year = {2020},
  month = {may},
  pages={174}
}

@article{PhysRevLett.129.132501,
  title = {Charge Radii of $^{55,56}\mathrm{Ni}$ Reveal a Surprisingly Similar Behavior at $N=28$ in Ca and Ni Isotopes},
  author = {Sommer, Felix and others.},
  journal = {Phys. Rev. Lett.},
  volume = {129},
  issue = {13},
  pages = {132501},
  numpages = {8},
  year = {2022},
  month = {Sep},
  publisher = {American Physical Society},
  doi = {10.1103/PhysRevLett.129.132501},
  url = {https://link.aps.org/doi/10.1103/PhysRevLett.129.132501}
}

@article{PhysRevLett.106.192501,
  title = {\textit{Ab Initio} Calculation of the Hoyle State},
  author = {Epelbaum, Evgeny and Krebs, Hermann and Lee, Dean and Mei\ss{}ner, Ulf-G.},
  journal = {Phys. Rev. Lett.},
  volume = {106},
  issue = {19},
  pages = {192501},
  numpages = {4},
  year = {2011},
  month = {May},
  publisher = {American Physical Society},
  doi = {10.1103/PhysRevLett.106.192501},
  url = {https://link.aps.org/doi/10.1103/PhysRevLett.106.192501}
}

@article{Elhatisari2015,
  title = {\textit{Ab initio} alpha–alpha scattering},
  volume = {528},
  ISSN = {1476-4687},
  url = {http://dx.doi.org/10.1038/nature16067},
  DOI = {10.1038/nature16067},
  number = {7580},
  journal = {Nature},
  publisher = {Springer Science and Business Media LLC},
  author = {Elhatisari,  Serdar and Lee,  Dean and Rupak,  Gautam and Epelbaum,  Evgeny and Krebs,  Hermann and L\"{a}hde,  Timo A. and Luu,  Thomas and Meißner,  Ulf-G.},
  year = {2015},
  month = {dec},
  pages = {111–114}
}

@article{Elhatisari2024,
  title = {Wavefunction matching for solving quantum many-body problems},
  volume = {630},
  ISSN = {1476-4687},
  url = {http://dx.doi.org/10.1038/s41586-024-07422-z},
  DOI = {10.1038/s41586-024-07422-z},
  number = {8015},
  journal = {Nature},
  publisher = {Springer Science and Business Media LLC},
  author = {Elhatisari,  Serdar and Bovermann,  Lukas and Ma,  Yuan-Zhuo and others.},
  year = {2024},
  month = {may},
  pages = {59–63}
}

@article{Gandolfi2020,
  title = {Atomic Nuclei From Quantum Monte Carlo Calculations With Chiral EFT Interactions},
  volume = {8},
  ISSN = {2296-424X},
  url = {http://dx.doi.org/10.3389/fphy.2020.00117},
  DOI = {10.3389/fphy.2020.00117},
  journal = {Front. Phys.},
  publisher = {Frontiers Media SA},
  author = {Gandolfi,  Stefano and Lonardoni,  Diego and Lovato,  Alessandro and Piarulli,  Maria},
  year = {2020},
  month = {apr},
  pages={117}
}

@article{PhysRevC.94.054007,
  title = {Local chiral potentials with $\mathrm{\ensuremath{\Delta}}$-intermediate states and the structure of light nuclei},
  author = {Piarulli, M. and Girlanda, L. and Schiavilla, R. and Kievsky, A. and Lovato, A. and Marcucci, L. E. and Pieper, Steven C. and Viviani, M. and Wiringa, R. B.},
  journal = {Phys. Rev. C},
  volume = {94},
  issue = {5},
  pages = {054007},
  numpages = {12},
  year = {2016},
  month = {Nov},
  publisher = {American Physical Society},
  doi = {10.1103/PhysRevC.94.054007},
  url = {https://link.aps.org/doi/10.1103/PhysRevC.94.054007}
}

@article{PhysRevLett.120.052503,
  title = {Light-Nuclei Spectra from Chiral Dynamics},
  author = {Piarulli, M. and Baroni, A. and Girlanda, L. and Kievsky, A. and Lovato, A. and Lusk, Ewing and Marcucci, L. E. and Pieper, Steven C. and Schiavilla, R. and Viviani, M. and Wiringa, R. B.},
  journal = {Phys. Rev. Lett.},
  volume = {120},
  issue = {5},
  pages = {052503},
  numpages = {6},
  year = {2018},
  month = {Feb},
  publisher = {American Physical Society},
  doi = {10.1103/PhysRevLett.120.052503},
  url = {https://link.aps.org/doi/10.1103/PhysRevLett.120.052503}
}

@article{PhysRevLett.133.212501,
  title = {Quantum Monte Carlo Calculations of Magnetic Form Factors in Light Nuclei},
  author = {Chambers-Wall, G. and Gnech, A. and King, G. B. and Pastore, S. and Piarulli, M. and Schiavilla, R. and Wiringa, R. B.},
  journal = {Phys. Rev. Lett.},
  volume = {133},
  issue = {21},
  pages = {212501},
  numpages = {8},
  year = {2024},
  month = {Nov},
  publisher = {American Physical Society},
  doi = {10.1103/PhysRevLett.133.212501},
  url = {https://link.aps.org/doi/10.1103/PhysRevLett.133.212501}
}

@article{PhysRevC.102.025501,
  title = {Chiral effective field theory calculations of weak transitions in light nuclei},
  author = {King, G. B. and Andreoli, L. and Pastore, S. and Piarulli, M. and Schiavilla, R. and Wiringa, R. B. and Carlson, J. and Gandolfi, S.},
  journal = {Phys. Rev. C},
  volume = {102},
  issue = {2},
  pages = {025501},
  numpages = {13},
  year = {2020},
  month = {Aug},
  publisher = {American Physical Society},
  doi = {10.1103/PhysRevC.102.025501},
  url = {https://link.aps.org/doi/10.1103/PhysRevC.102.025501}
}

@article{PhysRevC.107.014314,
  title = {Densities and momentum distributions in $A\ensuremath{\le}12$ nuclei from chiral effective field theory interactions},
  author = {Piarulli, M. and Pastore, S. and Wiringa, R. B. and Brusilow, S. and Lim, R.},
  journal = {Phys. Rev. C},
  volume = {107},
  issue = {1},
  pages = {014314},
  numpages = {11},
  year = {2023},
  month = {Jan},
  publisher = {American Physical Society},
  doi = {10.1103/PhysRevC.107.014314},
  url = {https://link.aps.org/doi/10.1103/PhysRevC.107.014314}
}

@article{PhysRevC.109.034005,
  title = {Maximally local two-nucleon interactions at ${\mathrm{N}}^{3}\mathrm{LO}$ in $\mathrm{\ensuremath{\Delta}}$-less chiral effective field theory},
  author = {Somasundaram, R. and Lynn, J. E. and Huth, L. and Schwenk, A. and Tews, I.},
  journal = {Phys. Rev. C},
  volume = {109},
  issue = {3},
  pages = {034005},
  numpages = {22},
  year = {2024},
  month = {Mar},
  publisher = {American Physical Society},
  doi = {10.1103/PhysRevC.109.034005},
  url = {https://link.aps.org/doi/10.1103/PhysRevC.109.034005}
}

@article{PhysRevLett.119.222505,
  title = {\textit{Ab initio} Calculations of the Isotopic Dependence of Nuclear Clustering},
  author = {Elhatisari, Serdar and Epelbaum, Evgeny and Krebs, Hermann and L\"ahde, Timo A. and Lee, Dean and Li, Ning and Lu, Bing-nan and Mei\ss{}ner, Ulf-G. and Rupak, Gautam},
  journal = {Phys. Rev. Lett.},
  volume = {119},
  issue = {22},
  pages = {222505},
  numpages = {6},
  year = {2017},
  month = {Dec},
  publisher = {American Physical Society},
  doi = {10.1103/PhysRevLett.119.222505},
  url = {https://link.aps.org/doi/10.1103/PhysRevLett.119.222505}
}

@article{Lu2019,
  title = {Essential elements for nuclear binding},
  volume = {797},
  ISSN = {0370-2693},
  url = {http://dx.doi.org/10.1016/j.physletb.2019.134863},
  DOI = {10.1016/j.physletb.2019.134863},
  journal = {Phys. Lett. B},
  publisher = {Elsevier BV},
  author = {Lu,  Bing-Nan and Li,  Ning and Elhatisari,  Serdar and Lee,  Dean and Epelbaum,  Evgeny and Meißner,  Ulf-G.},
  year = {2019},
  month = {oct},
  pages = {134863}
}

@article{Hagen2015,
  title = {Neutron and weak-charge distributions of the $^{48}$Ca nucleus},
  volume = {12},
  ISSN = {1745-2481},
  url = {http://dx.doi.org/10.1038/nphys3529},
  DOI = {10.1038/nphys3529},
  number = {2},
  journal = {Nat. Phys.},
  publisher = {Springer Science and Business Media LLC},
  author = {Hagen,  G. and Ekstr\"{o}m,  A. and Forssén,  C. and Jansen,  G. R. and others.},
  year = {2015},
  month = {nov},
  pages = {186–190}
}

@article{PhysRevC.62.014001,
  title = {Quantum Monte Carlo calculations of $A=8$ nuclei},
  author = {Wiringa, R. B. and Pieper, Steven C. and Carlson, J. and Pandharipande, V. R.},
  journal = {Phys. Rev. C},
  volume = {62},
  issue = {1},
  pages = {014001},
  numpages = {23},
  year = {2000},
  month = {Jun},
  publisher = {American Physical Society},
  doi = {10.1103/PhysRevC.62.014001},
  url = {https://link.aps.org/doi/10.1103/PhysRevC.62.014001}
}

@article{PhysRevC.89.024305,
  title = {Nucleon and nucleon-pair momentum distributions in $A\ensuremath{\le}12$ nuclei},
  author = {Wiringa, R. B. and Schiavilla, R. and Pieper, Steven C. and Carlson, J.},
  journal = {Phys. Rev. C},
  volume = {89},
  issue = {2},
  pages = {024305},
  numpages = {9},
  year = {2014},
  month = {Feb},
  publisher = {American Physical Society},
  doi = {10.1103/PhysRevC.89.024305},
  url = {https://link.aps.org/doi/10.1103/PhysRevC.89.024305}
}

@article{PhysRevC.110.054316,
  title = {Magnetic structure of $A\ensuremath{\le}10$ nuclei using the Norfolk nuclear models with quantum Monte Carlo methods},
  author = {Chambers-Wall, G. and Gnech, A. and King, G. B. and Pastore, S. and Piarulli, M. and Schiavilla, R. and Wiringa, R. B.},
  journal = {Phys. Rev. C},
  volume = {110},
  issue = {5},
  pages = {054316},
  numpages = {27},
  year = {2024},
  month = {Nov},
  publisher = {American Physical Society},
  doi = {10.1103/PhysRevC.110.054316},
  url = {https://link.aps.org/doi/10.1103/PhysRevC.110.054316}
}

@article{Barrett2013,
  title = {\textit{Ab initio} no core shell model},
  volume = {69},
  ISSN = {0146-6410},
  url = {http://dx.doi.org/10.1016/j.ppnp.2012.10.003},
  DOI = {10.1016/j.ppnp.2012.10.003},
  journal = {Prog. Part. Nucl. Phys.},
  publisher = {Elsevier BV},
  author = {Barrett,  Bruce R. and Navrátil,  Petr and Vary,  James P.},
  year = {2013},
  month = {mar},
  pages = {131–181}
}

@article{GarciaRuiz2016,
  title = {Unexpectedly large charge radii of neutron-rich calcium isotopes},
  volume = {12},
  ISSN = {1745-2481},
  url = {http://dx.doi.org/10.1038/nphys3645},
  DOI = {10.1038/nphys3645},
  number = {6},
  journal = {Nat. Phys.},
  publisher = {Springer Science and Business Media LLC},
  author = {Garcia Ruiz,  R. F. and others.},
  year = {2016},
  month = {feb},
  pages = {594–598}
}

@article{PhysRevC.101.014318,
  title = {Novel chiral Hamiltonian and observables in light and medium-mass nuclei},
  author = {Som\`a, V. and Navr\'atil, P. and Raimondi, F. and Barbieri, C. and Duguet, T.},
  journal = {Phys. Rev. C},
  volume = {101},
  issue = {1},
  pages = {014318},
  numpages = {19},
  year = {2020},
  month = {Jan},
  publisher = {American Physical Society},
  doi = {10.1103/PhysRevC.101.014318},
  url = {https://link.aps.org/doi/10.1103/PhysRevC.101.014318}
}

@article{PhysRevLett.109.052501,
  title = {Medium-Mass Nuclei with Normal-Ordered Chiral $NN\mathbf{+}3N$ Interactions},
  author = {Roth, Robert and Binder, Sven and Vobig, Klaus and Calci, Angelo and Langhammer, Joachim and Navr\'atil, Petr},
  journal = {Phys. Rev. Lett.},
  volume = {109},
  issue = {5},
  pages = {052501},
  numpages = {5},
  year = {2012},
  month = {Jul},
  publisher = {American Physical Society},
  doi = {10.1103/PhysRevLett.109.052501},
  url = {https://link.aps.org/doi/10.1103/PhysRevLett.109.052501}
}

@article{PhysRevLett.106.222502,
  title = {In-Medium Similarity Renormalization Group For Nuclei},
  author = {Tsukiyama, K. and Bogner, S. K. and Schwenk, A.},
  journal = {Phys. Rev. Lett.},
  volume = {106},
  issue = {22},
  pages = {222502},
  numpages = {4},
  year = {2011},
  month = {Jun},
  publisher = {American Physical Society},
  doi = {10.1103/PhysRevLett.106.222502},
  url = {https://link.aps.org/doi/10.1103/PhysRevLett.106.222502}
}

@Article{Epelbaum2015,
author={Epelbaum, E.
and Krebs, H.
and Mei{\ss}ner, U.-G.},
title={Improved chiral nucleon-nucleon potential up to next-to-next-to-next-to-leading order},
journal={Eur. Phys. J. A},
year={2015},
month={May},
day={12},
volume={51},
number={5},
pages={53},
issn={1434-601X},
doi={10.1140/epja/i2015-15053-8},
url={https://doi.org/10.1140/epja/i2015-15053-8}
}

@article{PhysRevC.83.031301,
  title = {Improved nuclear matter calculations from chiral low-momentum interactions},
  author = {Hebeler, K. and Bogner, S. K. and Furnstahl, R. J. and Nogga, A. and Schwenk, A.},
  journal = {Phys. Rev. C},
  volume = {83},
  issue = {3},
  pages = {031301},
  numpages = {5},
  year = {2011},
  month = {Mar},
  publisher = {American Physical Society},
  doi = {10.1103/PhysRevC.83.031301},
  url = {https://link.aps.org/doi/10.1103/PhysRevC.83.031301}
}

@article{PhysRevC.102.044333,
  title = {Reexamining the relation between the binding energy of finite nuclei and the equation of state of infinite nuclear matter},
  author = {Atkinson, M. C. and Dickhoff, W. H. and Piarulli, M. and Rios, A. and Wiringa, R. B.},
  journal = {Phys. Rev. C},
  volume = {102},
  issue = {4},
  pages = {044333},
  numpages = {13},
  year = {2020},
  month = {Oct},
  publisher = {American Physical Society},
  doi = {10.1103/PhysRevC.102.044333},
  url = {https://link.aps.org/doi/10.1103/PhysRevC.102.044333}
}

@article{PhysRevC.102.034313,
  title = {Exploring the relationship between nuclear matter and finite nuclei with chiral two- and three-nucleon forces},
  author = {Sammarruca, Francesca and Millerson, Randy},
  journal = {Phys. Rev. C},
  volume = {102},
  issue = {3},
  pages = {034313},
  numpages = {8},
  year = {2020},
  month = {Sep},
  publisher = {American Physical Society},
  doi = {10.1103/PhysRevC.102.034313},
  url = {https://link.aps.org/doi/10.1103/PhysRevC.102.034313}
}

@article{HUTHER2020135651,
title = {Family of chiral two- plus three-nucleon interactions for accurate nuclear structure studies},
journal = {Phys. Lett. B},
volume = {808},
pages = {135651},
year = {2020},
issn = {0370-2693},
doi = {https://doi.org/10.1016/j.physletb.2020.135651},
url = {https://www.sciencedirect.com/science/article/pii/S0370269320304548},
author = {Thomas Hüther and Klaus Vobig and Kai Hebeler and Ruprecht Machleidt and Robert Roth}
}

@Article{Miyagi2023,
author={Miyagi, Takayuki},
title={NuHamil : A numerical code to generate nuclear two- and three-body matrix elements from chiral effective field theory},
journal={Eur. Phys. J. A},
year={2023},
month={Jul},
day={08},
volume={59},
number={7},
pages={150},
issn={1434-601X},
doi={10.1140/epja/s10050-023-01039-y},
url={https://doi.org/10.1140/epja/s10050-023-01039-y}
}

@article{PhysRevLett.120.122502,
  title = {Properties of Nuclei up to $A=16$ using Local Chiral Interactions},
  author = {Lonardoni, D. and Carlson, J. and Gandolfi, S. and Lynn, J. E. and Schmidt, K. E. and Schwenk, A. and Wang, X. B.},
  journal = {Phys. Rev. Lett.},
  volume = {120},
  issue = {12},
  pages = {122502},
  numpages = {6},
  year = {2018},
  month = {Mar},
  publisher = {American Physical Society},
  doi = {10.1103/PhysRevLett.120.122502},
  url = {https://link.aps.org/doi/10.1103/PhysRevLett.120.122502}
}

@article{PhysRevC.109.064316,
  title = {\textit{Ab initio} calculations with a new local chiral ${\mathrm{N}}^{3}\mathrm{LO}$ nucleon-nucleon force},
  author = {Wang, P. Y. and Li, J. G. and Zhang, S. and Yuan, Q. and Xie, M. R. and Zuo, W.},
  journal = {Phys. Rev. C},
  volume = {109},
  issue = {6},
  pages = {064316},
  numpages = {8},
  year = {2024},
  month = {Jun},
  publisher = {American Physical Society},
  doi = {10.1103/PhysRevC.109.064316},
  url = {https://link.aps.org/doi/10.1103/PhysRevC.109.064316}
}

@article{PhysRevLett.133.072502,
  title = {Direct Observation of Competing $M1$ and $M3$ Transitions in $^{10}\mathrm{B}$},
  author = {Ku\ifmmode \mbox{\c{s}}\else \c{s}\fi{}o\ifmmode \breve{g}\else \u{g}\fi{}lu, A. and others},
  journal = {Phys. Rev. Lett.},
  volume = {133},
  issue = {7},
  pages = {072502},
  numpages = {7},
  year = {2024},
  month = {Aug},
  publisher = {American Physical Society},
  doi = {10.1103/PhysRevLett.133.072502},
  url = {https://link.aps.org/doi/10.1103/PhysRevLett.133.072502}
}

@article{SHIMIZU2019372,
title = {Thick-restart block Lanczos method for large-scale shell-model calculations},
journal = {Comput. Phys. Commun.},
volume = {244},
pages = {372-384},
year = {2019},
issn = {0010-4655},
doi = {https://doi.org/10.1016/j.cpc.2019.06.011},
url = {https://www.sciencedirect.com/science/article/pii/S0010465519301985},
author = {Noritaka Shimizu and Takahiro Mizusaki and Yutaka Utsuno and Yusuke Tsunoda}
}

@article{PhysRevC.92.034331,
  title = {Magnus expansion and in-medium similarity renormalization group},
  author = {Morris, T. D. and Parzuchowski, N. M. and Bogner, S. K.},
  journal = {Phys. Rev. C},
  volume = {92},
  issue = {3},
  pages = {034331},
  numpages = {12},
  year = {2015},
  month = {Sep},
  publisher = {American Physical Society},
  doi = {10.1103/PhysRevC.92.034331},
  url = {https://link.aps.org/doi/10.1103/PhysRevC.92.034331}
}

@article{PhysRevLett.118.262501,
  title = {First Elastic Electron Scattering from $^{132}\mathrm{Xe}$ at the SCRIT Facility},
  author = {Tsukada, K. and others},
  journal = {Phys. Rev. Lett.},
  volume = {118},
  issue = {26},
  pages = {262501},
  numpages = {5},
  year = {2017},
  month = {Jun},
  publisher = {American Physical Society},
  doi = {10.1103/PhysRevLett.118.262501},
  url = {https://link.aps.org/doi/10.1103/PhysRevLett.118.262501}
}

@article{10.1093/ptep/pts043,
    author = {Suda, Toshimi and others},
    title = {Nuclear physics at the SCRIT electron scattering facility},
    journal = {Prog. Theor. Exp. Phys.},
    volume = {2012},
    number = {1},
    pages = {03C008},
    year = {2012},
    month = {12},
    issn = {2050-3911},
    doi = {10.1093/ptep/pts043},
    url = {https://doi.org/10.1093/ptep/pts043}
}

@Misc{imsrgcode,
    author = {S. R. Stroberg},
    year={2024},
    howpublished = {\url{https://github.com/ragnarstroberg/imsrg}}
}

\end{document}